\documentclass[onecolumn,pre,amsmath,amssymb]{revtex4}
\usepackage{graphicx}% Include figure files
\usepackage[latin1]{inputenc}
\usepackage{dcolumn}% Align table columns on decimal point
\usepackage{bm}% bold math
\usepackage{epsfig}
\usepackage{color}
\usepackage{ulem}
\begin{document}

\title{Irreversibility of financial time series: a graph-theoretical approach}% Force line breaks with \\

\author{Ryan Flanagan and Lucas Lacasa}
\affiliation{School of Mathematical Sciences, Queen Mary University of London}%


\begin{abstract}
The relation between time series irreversibility and entropy production has been recently investigated in thermodynamic systems operating away from equilibrium. In this work we explore this concept in the context of financial time series. We make use of visibility algorithms to quantify in graph-theoretical terms time irreversibility of 35 financial indices evolving over the period 1998-2012. We show that this metric is complementary to standard measures based on volatility and exploit it to both classify periods of financial stress and to rank companies accordingly. We then validate this approach by finding that a projection in principal components space of financial years based on time irreversibility features clusters together periods of financial stress from stable periods. Relations between irreversibility, efficiency and predictability are briefly discussed.
\end{abstract}

 \maketitle

\section{Introduction}
\indent The quantitative analysis of financial time series \cite{book_financial} is a classical field in econometrics that has received in the last decades valuable inputs from statistical physics, nonlinear dynamics and complex systems communities (see \cite{seminal1, seminal2, book_econo} and references therein for seminal contributions). In particular, the presence of long-range dependence, detected via multifractal measures has been used to quantify the level of development of a given market \cite{intro_1}. This approach was subsequently extended to address the problem of quantifying the degree of market inefficiency \cite{intro_12, intro_13, intro_14, intro_15, intro_16}. Here the adjective \textit{efficient} refers to a market who is capable of integrating, at any given time, all available information in the price of an asset (the so-called weak-form of the Efficient Market Hypothesis). In such a situation, assets should follow a martingale process in which each price change is unmodified by its predecessor, and the possibility of arbitrage would be impossible. Deviations from this ideal behavior are thus characterizing an inefficient system. In another approach, using the notion of forbidden patterns and permutation entropy, concepts arising in nonlinear dynamics, it was shown that one can also evaluate the degree of inefficiency of a market \cite{intro_2}. These approaches are indeed relevant from an applied perspective, as it has been shown that some links exist between the degree of inefficiency of a market and its predictability \cite{predict}. 

\noindent Another relevant property in the context of financial systems is that of time reversibility, i.e. the degree of dynamical invariance under time reversal. Statistical time reversibility refers to the situation where the statistical properties of a certain process are invariant under time reversal \cite{weiss}, this being a more natural concept to explore in noisy series than strict reversibility, and intuitively measures our capacity to detect the correct arrow of time in erratically evolving dynamics. This purely statistical concept has been recently found to have deep links with the physics of information, as physical systems operating away from equilibrium are shown to produce entropy at a rate which is proportional to a suitable measure of time irreversibility of adequate physical observables \cite{edgar,edgar2}. This concept has received little attention in the financial realm (see however some initial investigations on this matter \cite{fin0,fin1,fin2,fin3,fin4}), perhaps due to the fact that financial time series are usually non-stationary \cite{book_financial}, and both the concepts of time irreversibility and its associated entropy production are not well defined in that case. Interestingly, a recent approach to time series analysis proceeds by transforming a series into a graph and exploring the topological properties of the graph-theoretical representation in order to describe the underlying dynamical process. Among other methods, the family of visibility algorithms \cite{pnas,pre,epjb} have been recently shown to be well suited to capture time series irreversibility in both stationary and non-stationary systems \cite{RW_visibility}, thus enabling the study of such property in the financial realm. Accordingly, here we propose to apply the concept of graph-theoretical time series irreversibility in the context of financial time series. We first make use of the visibility algorithms to construct graph-theoretical representations of the stock prices of 35 companies from the New York Stock Exchange in the period 1998-2012. We then estimate time irreversibility in these representations through the Kullback-Leibler divergence of the {\it in} and {\it out} degree distributions. After checking that this measure is indeed genuine and not correlated to volatility, we show that all the companies under study are irreversible, but their degree of irreversibility varies across companies and fluctuates over time. The variance across companies allows us to rank companies, and the collective time fluctuations are finally used to provide a classification of financial periods.

\section{Methods}

\noindent {\bf Irreversibility and entropy production. }\\
A dynamical process is said to be time reversible, if any two time series ${\cal S}=\{x_1,x_2,\dots,x_n\}$ and ${\cal S}^-=\{x_{-1},x_{-2},\dots,x_{-n}\}$ (where $n$ denotes time) generated by this process have asymptotically the same joint distribution \cite{weiss}. In the concrete case where the process is stationary, the time series $\{x_{-1},x_{-2},\dots,x_{-n}\}$ and $\{x_{-1+m},x_{-2+m},\dots,x_{-n+m}\}$ have the same joint distributions $\forall m$, so for the particular choice $m=n+1$, the definition of time reversibility reduces to the equivalence of statistics between the forward and backward process:
a stationary time series is thus time reversible if a series $\{x_1,x_2,\dots,x_n\}$ and its reverse $\{x_n,\dots,x_2,x_1\}$ are equally likely to occur. That is to say, the joint distributions $p(x_1,x_2,\dots,x_n)$ and $q(x_n,\dots,x_2,x_1)$ coincide for reversible processes. If $p\neq q$ we say that the process is (statistically) time irreversible. Examples of reversible processes include white noise, linearly correlated Gaussian processes, and thermodynamic systems close to equilibrium, whereas examples of irreversible processes include typically chaotic dissipative processes, nonlinear stochastic processes and processes with memory, operating away from thermodynamic equilibrium. Intuitively speaking, it is obvious that a physical process is time irreversible if one is able to clearly know which is the "correct" arrow of time when the process is observed forward and backwards in time. Similarly, one should be able to make easier predictions on irreversible processes, where the arrow of time is playing a role, than on reversible ones. For instance, consider a chaotic system initially evolving in a high-dimensional phase space. If the system is measure-preserving (e.g. Hamiltonian dynamics), then it is more likely to be statistically time reversible \cite{epjb}, whereas if the process is dissipative it is usually irreversible. In the latter case, trajectories tend to converge to a low-dimensional attractor. In some sense, prediction of future states is easier in this case, if only because the system is confined to a manifold of lower dimension and the number of effective degrees of freedom required to describe the state of the system is smaller (less uncertainty).
This somewhat speculative relation between irreversibility and predictability is an additional motivation for the study of the statistical irreversibility in financial series. 

\noindent In the literature several possible ways of quantifying (statistical) time irreversibility have been proposed, usually exploring the asymmetries between the statistics (i.e., the probability distributions) arising in the forward and backward process. If series are discrete (such as Markov chains), estimation of probability distributions is straightforward. However for real-valued time series, this estimation requires to perform a time series {\it symbolization} as a pre-processing:
 defining a certain alphabet of symbols, one can transform a real-valued series into a discrete sequence of symbols (i.e., one has to make a partition of the original phase space). Whereas this is common practice, we must recall that both the particular partition and the number of symbols are free parameters that need to be tuned specifically for each system, and might introduce ambiguities or undesired partition-dependences in the subsequent results \cite{multivariate} (it is very easy to see that unlucky partitions can easily translate periodic series into constant series, chaotic series into periodic ones, etc). As we will show later, by using visibility algorithms we circumvent the ambiguities associated to the symbolization problem. 
 
Amongst other descriptors, we advocate that the Kullback-Leibler divergence (KLD) between the statistics (distributions) of the (appropriately symbolized) time series of the forward and backward process is indeed an interesting choice. We recall that if $p$ and $q$ are discrete distributions with domain ${\cal X}$, then the Kullback-Leibler divergence $D_{\text{kld}}(p|q)$ is defined as
\begin{equation}
D_{\text{kld}}(p|q)=\sum_{{\bf x}\in \cal X} p({\bf x})\log \frac{p({\bf x})}{q({\bf x})}
\end{equation}
This is a semi-distance (i.e., non-symmetric) which is null if and only if $p=q$ and positive otherwise.  First, this measure is relevant on information-theoretic grounds: the probability of failing a hypothesis test and misleadingly confounding $p$ and $q$ decreases as $\exp(-D_{\text{kld}}(p|q))$ according to Chernof-Stein lemma. More importantly, if the process is stationary and the series $\{x_1,x_2,\dots,x_n\}$ is understood as the evolution of a certain system's observable (such as the position of a Brownian particle fluctuating in a certain environment), then it can be proved that $D_{\text{kld}}(p|q)$ (where $p$ and $q$ describe the statistics of the {\it forward} and {\it backward} process) gives a lower bound to the physical dissipation and entropy production of the particle \cite{edgar2, edgar}: for $D\to 0$ the system is in thermodynamic equilibrium, whereas for $D>0$ the system is evolving away from equilibrium. This bound is tight when the measured process that gives rise to $p$ encloses all information (i.e., in the limit ${\bf x}=\lim_{p\to \infty} (x_1,x_2,\dots,x_p)$). Hence the Kullback-Leibler divergence between forward and backward statistics provides physical information on the state of a system even if one ignores the microscopic details (i.e., the dynamics) of this system. The concept of time series irreversibility, via KLD, thus reveals a deep link between information theory and statistical physics. In this work we aim at exploiting this link in the realm of financial systems. If one understands financial indicators evolving over time as the physical observables of the underlying (statistical-mechanical) system, then the irreversibility of these observables provides a lower bound on the system's dissipation. One is thus entitled to consider the following questions: Is the financial system `in thermal equilibrium'? How do financial crisis and other major perturbations drive the financial system `away from equilibrium'? Which companies are evolving closer to equilibrium (and therefore are producing less entropy)? What is the relation between irreversibility and predictability, in the framework of financial series? 

\noindent It is important to stress at this point that financial time series are usually non-stationary. This is in principle a fundamental drawback, as to the best of our knowledge no rigorous theory has been advanced so far linking time series irreversibility and entropy production in the non-stationary case. As a matter of fact, according to the original definition, non-stationary series are infinitely irreversible, so the quantification of how irreversible a non-stationary time series seems to be an ill-defined problem to begin with. Again, here we circumvent this problem by using the so-called visibility algorithms, a family of methods to make time series analysis in graph space that have been shown recently to be able to quantify different degrees of irreversibility in both stationary and non-stationary processes \cite{RW_visibility}. 

\begin{figure}
\centering
\includegraphics[width=0.65\columnwidth]{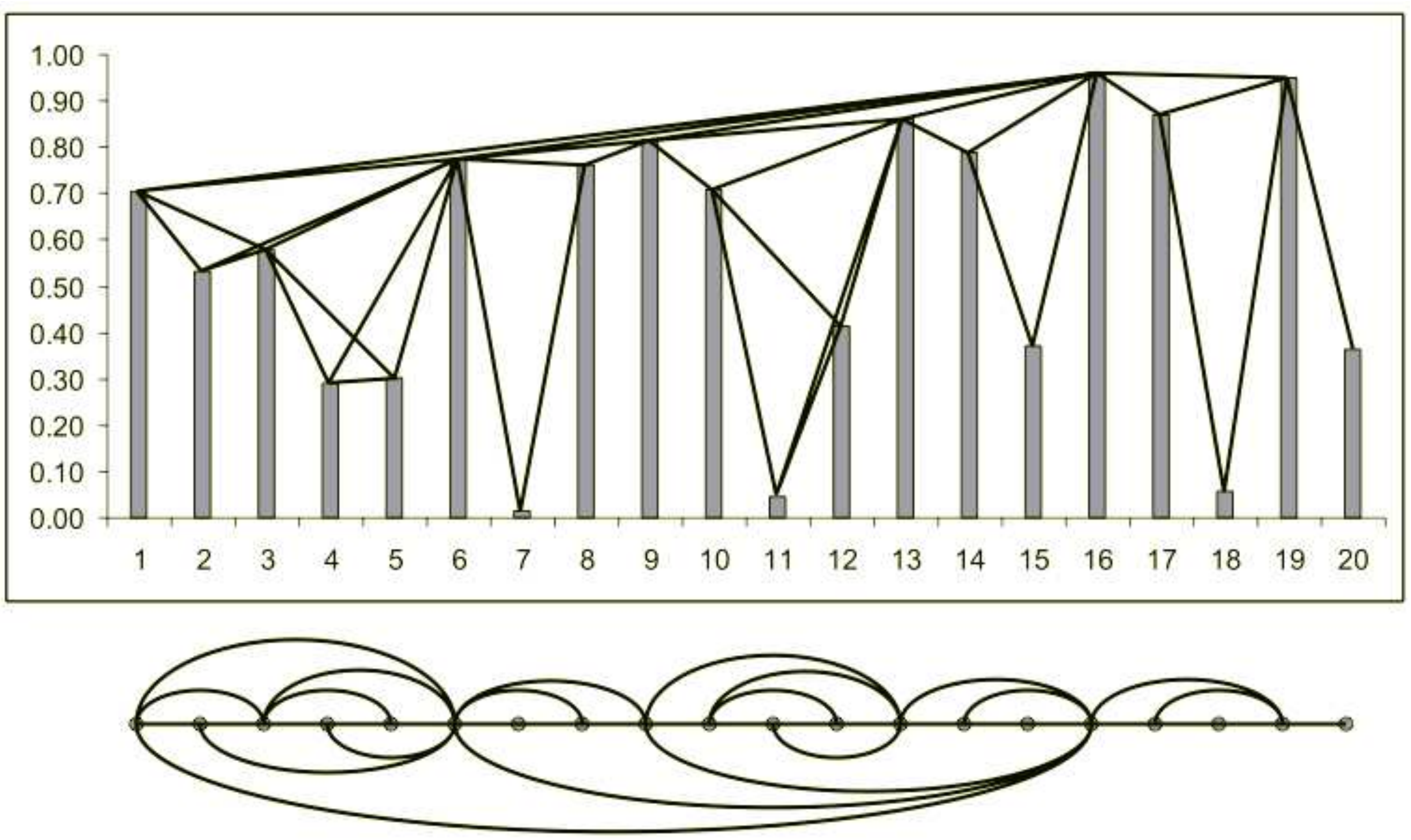}
\caption{Sample time series of 20 values. The visibility graph of this series is represented below (see the text for details.}
\label{Fig_VG}
\end{figure}

\noindent {\bf Visibility algorithms}\\ 
Visibility algorithms \cite{pnas, pre, epjb} are a family of methods to map time series into graphs, in order to explore the structure of time series (and the dynamics underneath) using graph theory. Let ${\cal S}=\{x(t)\}_{t=1}^T$ be a real-valued time series of $T$ data. A so called natural visibility graph (VG) is a graph of $T$ nodes associated to $\cal S$, such that (i) every datum $x(i)$ in the series is mapped to a node $i$ in the graph (hence the graph nodes inherit a natural ordering), and (ii) two nodes $i$ and $j$ are connected by an edge if the associated data show mutual natural visibility. More precisely, if any other datum $x(k)$ where $i<k<j$ fulfils the following \textit{convexity} criterion:
$$x_k< x_i + \frac{k-i}{j-i}[x_j-x_i],\ \forall k: i<k<j$$
By construction, VGs are planar connected graphs, and this construction is invariant under a set of basic transformations in the series, including horizontal and vertical translations. An illustration of this method is shown in figure \ref{Fig_VG}, where we plot a time series of 20 data and its associated VG. An important property of these graphs is that they are well suited to investigate the properties of non-stationary signals. For instance, it was shown \cite{epl} that the degree distribution of VGs associated to series generated by a (non-stationary) fractional Brownian motion with Hurst exponent $H$ have a power-law tail with exponent $\gamma = 3-2H$. This analysis has been subsequently applied to finance \cite{VG_finance1, VG_finance4}, fluid dynamics \cite{VG_turbulence, VG_turbulence2}, or medical research \cite{VG_EEG} to cite a few. Other works applying VG to finance deal with properties such as spanning trees \cite{VG_finance2}, or community structure \cite{VG_finance3}. 

\noindent A so called horizontal visibility graph (HVG) is defined as a subgraph of the VG, obtained by restricting the visibility criterion and imposing horizontal visibility instead. In this case, two nodes $i$ and $j$ are connected by an edge in the HVG if any other datum $x(k)$ where $i<k<j$ fulfill the following \textit{ordering} criterion:
 $$x_k<\inf(x_i,x_j),\ \forall k: i<k<j$$
Such subgraph is indeed an outerplanar graph \cite{severini}. Several analytical properties of these family of graphs associated to different classes of dynamics have been analytically investigated in recent years \cite{nonlinearity, plosone, pre2013, jpa2014, jns}. 

\noindent Note that previous definitions generate undirected graphs. However, these can be made {\it directed} \cite{epjb} by assigning to the links the time arrow naturally induced by the node ordering. Accordingly, a link between $i$ and $j$ (where time ordering yields $i<j$), generates an {\it outgoing} link for $i$ and an {\it ingoing} link for $j$ in a directed version of a VG/HVG. The degree sequence of the VG/HVG (which assigns to each node its degree or number of edges) thus splits into an ingoing degree sequence $\{k_{in}(t)\}_{t=1}^T$, where $k_{in}(t)$ is the ingoing degree of node $i=t$, and an outgoing degree sequence. An important property at this point is that the ingoing and outgoing degree sequences are interchangeable under time series reversal. That is to say, if we define the time reversed series ${\cal S}^*=\{x_{T+1-t}\}_{t=1}^T$, then we have the following identities
\begin{equation}
\{k_{in}(t)\}[{\cal S}]=\{k_{out}(t)\}[{\cal S}^*]; \ \{k_{out}(t)\}[{\cal S}]=\{k_{in}(t)\}[{\cal S}^*],
\label{degseqrev}
\end{equation}
that is, the in (out) degree sequence of the VG/HVG associated to ${\cal S}$ is identical to the out (in) degree sequence of the VG/HVG associated to ${\cal S}^*$.
Now, one can define, from the ingoing and outgoing degree sequences, 
an ingoing degree distribution $P(k_{in})\equiv P_{in}(k)$ and an outgoing degree distribution $P(k_{out})\equiv P_{out}(k)$, and property (\ref{degseqrev}) is inherited in the distributions, such that
\begin{equation}
P_{in}(k)[{\cal S}]=P_{out}(k)[{\cal S}^*]; \ P_{out}(k)[{\cal S}]=P_{in}(k)[{\cal S}^*]
\label{degdisrev}
\end{equation}
In other words, the statistics of the forward and the backward process are encoded, in graph-space, in the {\it in} and {\it out} degree sequences.
Time series irreversibility can then be estimated via the Kullback-Leibler divergence between the {\it in} and {\it out} degree distributions \cite{epjb}. This graph-theoretical measure reads
\begin{equation}
I_{\text{HVG/VG}}({\bf x}):= D_{\text{kld}}(P_{\text{in}}||P_{\text{out}})
\label{Irr}
\end{equation}
%\noindent A multiplex version has also recently been proposed. A multivariate time series $\{\textbf{x}(t)\}_{t=1}^N, \ \textbf{x} \in \mathbb{R^d}$ is mapped into a (horizontal) visibility multiplex of $d$ layers, where layer $\alpha$ ($\alpha=1,\dots,d$) is the (horizontal) visibility graph of the time series associated to the $\alpha$ component of $\textbf{x}$. Directed visibility multiplexes trivially follow.
$I_{\text{HVG/VG}}$ tends to zero as series size increases for HVG/VG reversible processes, and converges to a positive, finite value for HVG/VG irreversible ones (note that we will always find $I_{\text{HVG/VG}}>0$ for finite processes, as even for reversible processes, finite size effects always prevent the statistics of forward and backward process to be identical). 

\noindent This latter measure was introduced in \cite{epjb} and has been extended and applied to several scenarios \cite{donner, suyal, Xie, donner1, Zou}. Interestingly, these methods do not require to perform any hoc time series symbolizations (as the series under study -degree sequences- are by construction discrete), and therefore are free from the ambiguities associated to the effect that different partitions and alphabets produce commented above. 

\noindent Very recently \cite{RW_visibility}, the performance of VG/HVG in assessing time series irreversibility in {\it non-stationary systems} was addressed theoretically. It was found that the invariance properties of VG and HVG under certain transformations enable them to explore time irreversibility in the same grounds for both stationary and non-stationary processes. Whereas non-stationary processes are infinitely irreversible according to standard definitions, it was shown that in VG/HVG space these measures are `renormalized', yielding finite irreversibility measures whose value quantify the onset of different aspects (memory effects, trends, etc). As such, unbiased random walks have vanishing irreversibility measures, as expected intuitively, whereas biased or non Markovian random walks are irreversible. As a result, VG/HVG seem to be a well-defined tool to assess irreversibility in financial series.

\begin{figure}
\centering
\includegraphics[width=0.65\columnwidth]{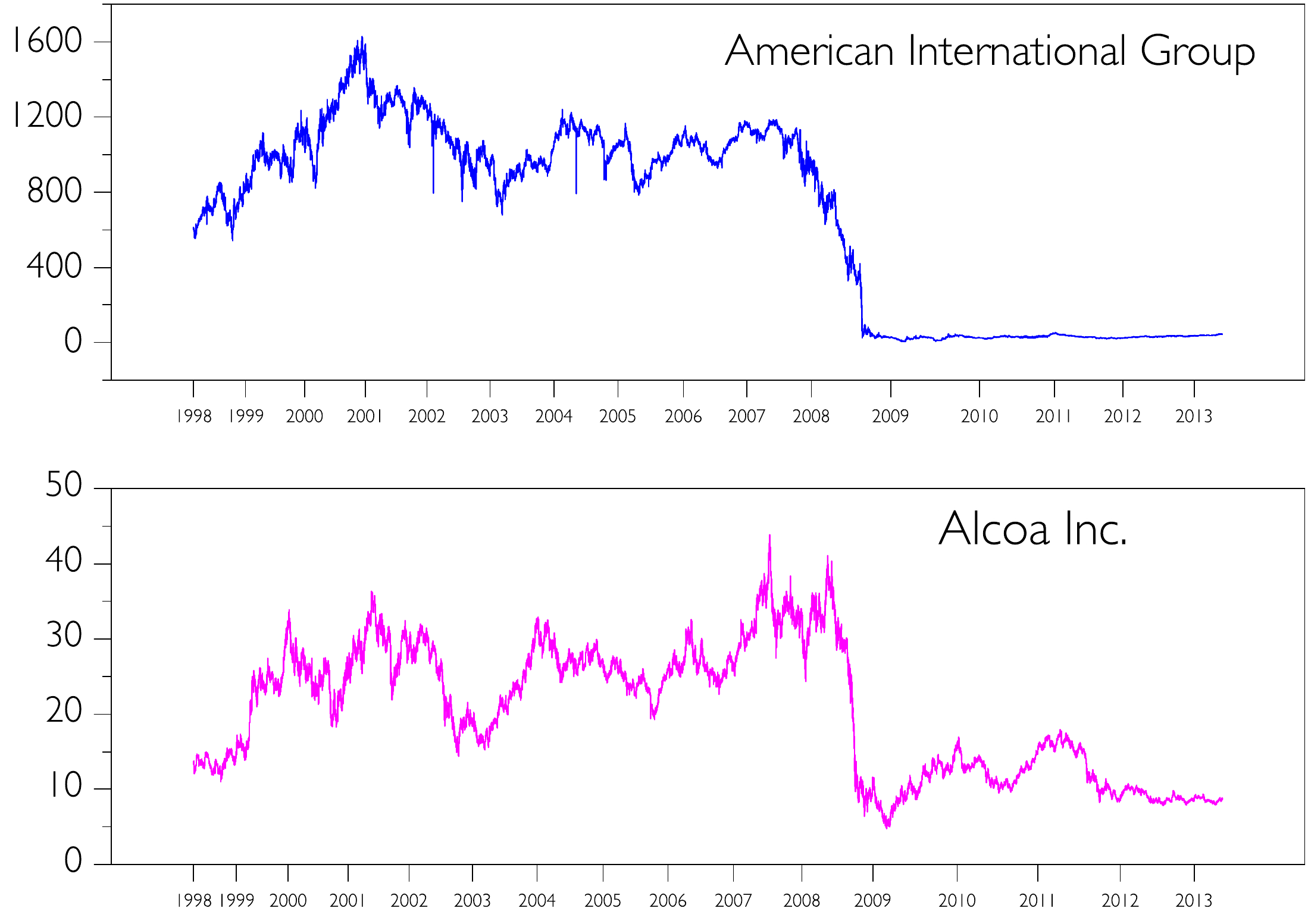}
\caption{Stock price series $x(t)$ (adjusted closing price) as a function of time, for two different companies: Alcoa Inc. and Bank of America. Note that time is not strictly equispaced as there are some missing data for each company.}
\label{example2}
\end{figure}

\section{Data and Results}
We have analyzed a dataset of financial stocks comprising stock evolution of 35 major American companies from the New York Stock Exchange (NYSE) and Nasdaq in the period 1998-2012, the majority of which belong to the Dow Jones Industrial Average (see table \ref{companies} for a list). NYSE is the largest and most liquid cash equities exchange in the world by market capitalization. It therefore represents an appropriate set of observables to study the underlying evolution of the financial system. The series have very high resolution (intraday resolution of approximately one data per minute, in 1998-2012), yielding ${\cal O}(2\cdot10^6)$ data per company, which allows us to make a fine-grained and statistically robust analysis, and to break down data into different periods without losing statistical accuracy. We use the adjusted closing prices for concreteness. For illustration purposes, in figure \ref{example2} we plot two sample series, representing the evolution of American International Group and Alcoa Inc. (note that series have been downsampled in this figure; each point is taken every 1000 time stamps). The dynamic range in this example differs greatly between these two companies, and this is heterogeneity actually extends to the rest of the companies under consideration.

\noindent {\bf Basic measures of Irreversibility} \\
In this section we explore and assess the irreversible character of financial data over the period 1998-2012. A priori, note that we can consider two different time series, namely the standard (non-stationary) price $x(t)$ and the log-returns $r(t)=\log(x(t)/x(t-1))$. This latter one is typically used in finance instead of $x(t)$, because it is believed that it is more stationary (a property of utmost importance for time series analysis \cite{book_financial}). Furthermore, because prices usually fluctuate by increasing or decreasing in terms of a percentage of the price, stochastic models of these price fluctuations acquire a more natural interpretation in logarithmic space. Historically it has been customary to use the rate of change $x(t+1)/x(t)$ as a random variable $\xi$. Thus $x(t+1)=\xi \cdot x(t)$ and this multiplicative process directly yields that $r(t)$ behaves as an i.i.d. random variable (white noise), which is stationary. 

\noindent As previously stated, visibility algorithms are well suited \cite{RW_visibility} to study the irreversibility in non-stationary processes, hence it is not necessary to work with log-returns and we can directly use the original price series $x(t)$. As the underlying dynamics of $x(t)$ are expected to follow a multiplicative rather than an additive stochastic process, we shall focus on VG rather than HVG according to previous theory \cite{RW_visibility}.

\begin{figure}
\centering
\includegraphics[width=0.65\columnwidth]{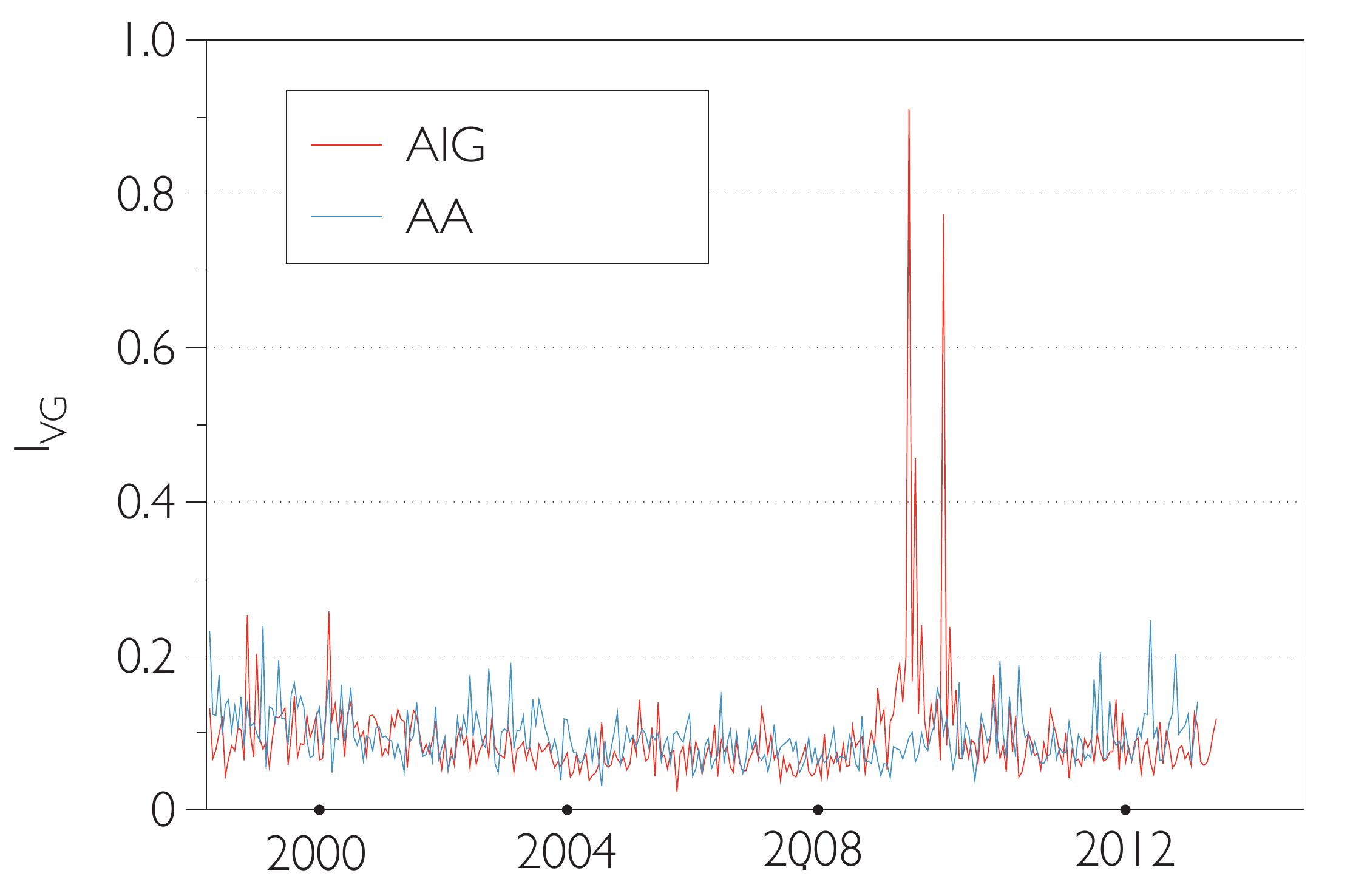}
\caption{(Color Online) VG Irreversibility measure as a function of time (each dot represents $I_{\text{VG}}$ associated to a non-overlapping time window of 5000 data points), for AIG (American International Group) and AA (Alcoa Inc.). The impact of the global financial crisis is only evident for AIG.}
\label{example}
\end{figure}

\noindent Our methodology is as follows: for each company $c$, we initially consider its price time series $x(t)$ all over the period 1998-2012. We then define a working time window of $n=5000$ data points, and divide our original time series $\{x(t)\}$ of $N$ data into a collection of $N/n$ non-overlapping  time series of $n$ data each. For each of these sub-series, we construct its associated VG, and compute the irreversibility measure $I_{\text{VG}}$ according to eq.\ref{Irr} (results using the $\ell_1$ norm are qualitatively similar), yielding a vector $I_{\text{VG}}^c({\bf w})$ for company $c$, ${\bf w}=(w_1,w_2,\dots, x_{N/n})$. 
As a technical remark, note that $D_{\text{kld}}(p|q)$ diverges if $p$ and $q$ have different supports (i.e., if $q(m)=0, p(m)\neq 0$ or $p(m)=0,\  q(m)\neq 0$ for some value $m$). In order to take appropriately weight this possibility while maintaining the irreversibility measure finite in pathological cases, a common procedure \cite{edgar} is to introduce a small bias that allows for the possibility of having a small uncertainty for every contribution. Here we introduce a bias of order ${\cal O}(1/n^2)$ where $n$ is the series size (i.e., we replace all vanishing frequencies with $1/n$, and we normalize the frequency histogram appropriately). 

\noindent We have thus computed $I_{\text{VG}}^c({\bf w})$ for all 35 companies in our dataset. All values obtained comply with the presence of VG irreversibility, as finite-size irreversibility values (for window size $n=5000$) are generally higher than those found for reversible null models (white noise, additive random walks) \cite{RW_visibility}. This was indeed expected, as multiplicative models are indeed known to display VG irreversibility \cite{RW_visibility}, and coincides with previous evidence \cite{fin0,fin1,fin2,fin3,fin4}. Interestingly however, this quantity is fluctuating over time, and there are periods where the reversiblity is comparable to additive random walks, thus we can state that in general stock prices are irreversible but periods of quasi-reversibility are not uncommon. Following the conceptual link between predictability and efficiency, one can infer that the market is more efficient when the stock price's reversibility approaches that found in reversible null models. In periods of financial stress the irreversibility increases and the system is less efficient, and therefore somehow more predictable. For illustration purposes, the evolution of irreversibility for the two companies shown in figure \ref{example} (American International Group (AIG) and Alcoa Inc. (AA)) is plotted in figure \ref{example}. One can immediately appreciate several differences among the two companies. For instance, whereas Alcoa Inc. (a metals technology corporation) seems to have a rather stable irreversibility over time, American International Group (a financial corporation) exhibits an abnormal irreversibility increase from year 2008, peaking in 2009. In principle one could ask whether there is a direct relation between the dynamic range of a certain series and its irreversibility. In the next section we rule out this possibility by observing a very small correlation between irreversibility and volatility.

\noindent The results for all companies are not shown but it is important to note that the amount of irreversibility is not indeed a stable quantity, neither intracompany (i.e., for the same company, over different periods), nor across companies. As pointed out above, intracompany irreversibility heterogeneity points out to the influence of exogenous factors, such as the impact of financially unstable periods. On the other hand, the fact that different companies have different irreversibility patterns is indicative that each observable of the financial system evolves over time in a different fashion, and thus this property can be used to rank companies accordingly. In the next sections we investigate these aspects.

\noindent {\bf Ranking companies.} \\
In order to quantify the net amount of irreversibility of a certain company, we introduce $\text{Score}[c]$, the score of a company $c$ as the average of the annualized irreversibility value
\begin{equation}
\text{Score}[c]=\frac{1}{15}\sum_{\text{year}=1998}^{2012}I_{\text{VG}}^c(\text{year})
\label{Z}
\end{equation}
This quantity averages the degree of irreversibility of a given company over large periods of time. According to the analogy between reversibility and entropy production, the larger the Score is, the more `away from equilibrium' the signal generated by $c$ is, thus producing larger amounts of entropy. This might be relevant from a financial perspective, as the larger Score of a company, the less efficient it is and thus more interesting from an investment viewpoint. A company ranking can be made accordingly. The first five companies of such ranking are depicted in table \ref{top5companies} (see table \ref{companies} for the rest). No obvious interpretation can be stated at this point, as one finds multinationals operating in different sectors (insurance, industry, health) in this top rank. 

\begin{figure}
\centering
\includegraphics[width=0.65\columnwidth]{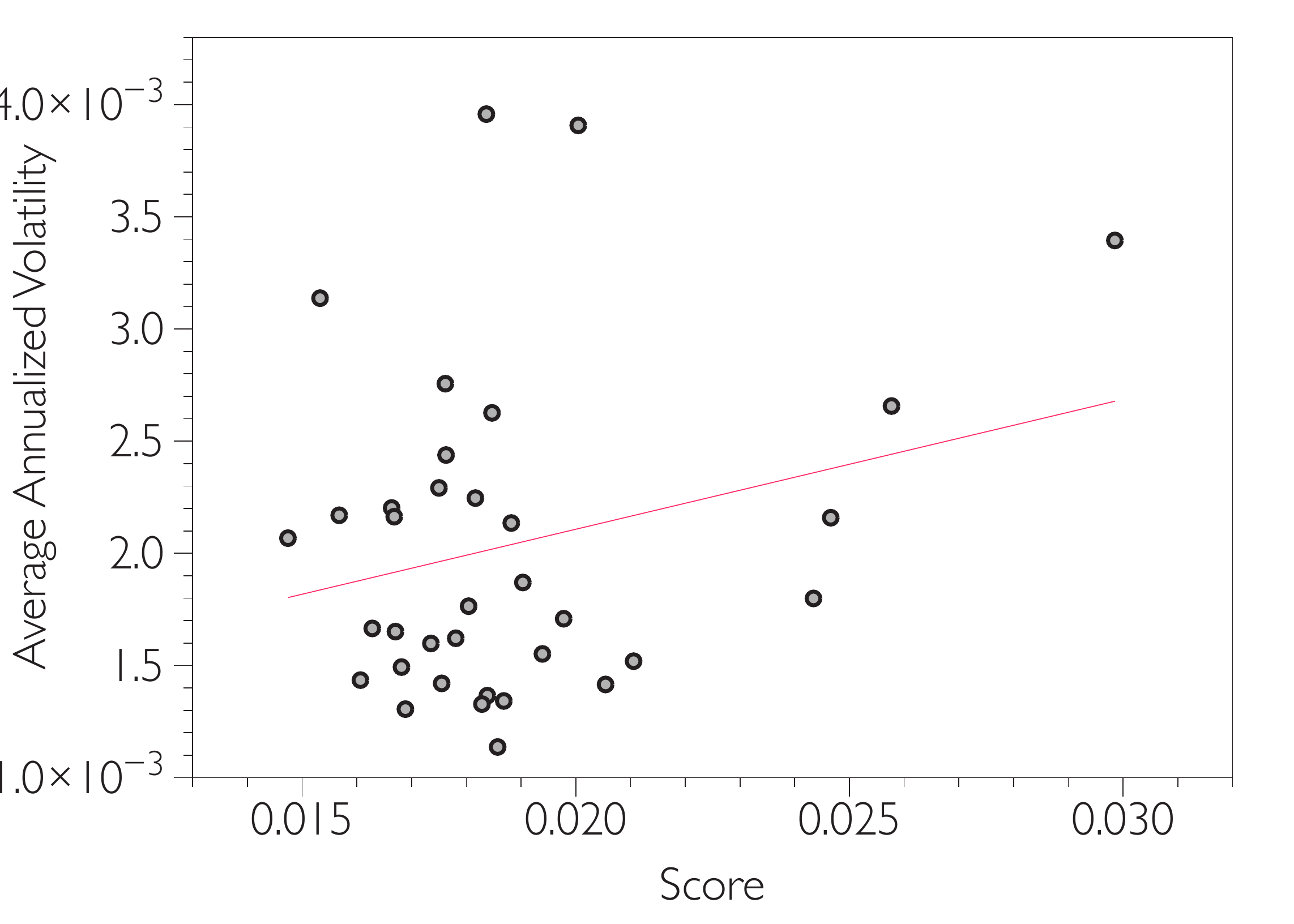}
\caption{Scatter plot of the Score against the Average Annualized Volatility of each of the 35 companies. The figure is highly scattered (the solid red line provides the best fitting to a linear dependence between both quantities, with Pearson's $r^2=0.06$). This indicates that the volatility and the irreversibility Score are not correlated, suggesting that the latter is a genuinely new and complementary measure that provides different information about the evolution and performance of a given company.}
\label{scores2}
\end{figure}

\begin{table}
\begin{ruledtabular}
\begin{tabular}{ccc}
{\bf Score Rank}&{\bf Acronym} &{\bf Name}\\
1&GM&General Motors Company\\
2&AIG&American International Group\\
3&TRV&The Travelers Companies, Inc.\\
4&AA&Alcoa Inc.\\
5&UNH&UnitedHealth Group Incorporated\\
\end{tabular}
\end{ruledtabular}
\caption{Top five companies according to the irreversibility score (eq. \ref{Z}) ranking.}
	\label{top5companies}
\end{table}
\noindent We now compare the new defined metric with standard financial metrics; we make use of the annualized volatility, which is commonly used to capture the dynamic range of financial data. We will define volatility as the standard deviation of the price log-returns over a year.  In figure \ref{scores2} we plot, for each company, the averaged annualized volatility (defined as the average of annualized volatilities over 1998-2012) against its Score. If both measures were correlated, we would expect that a smooth curve emerges in the scatter plot. The scatter, however, is large. The red line describes the best fitting of the data to a linear relation, with a poor Pearson's linear correlation $r^2=0.06$. We conclude that volatility and irreversibility are not correlated, making the latter a genuine and complementary metric.

\noindent To further assess the possibility that some companies may have suffered from large irreversibility only at sporadic occasions (which would yield a high irreversibility score even if the company were following a quasi-reversible evolution in most of the period), we also compute the irreversibility variance $\sigma^2(D_{\text{kld}})$ of a given company
\begin{equation}
\sigma^2(D_{\text{kld}})=\langle  I_{\text{VG}}^2 \rangle_{\text{years}} 
-\langle I_{\text{VG}} \rangle_{\text{years}}^2
\end{equation}

\noindent A company with low irreversibility variance corresponds to one whose evolution is relatively independent of the particular strength of exogenous variables; its response against external perturbations is relatively independent of the strength of the perturbation, and only depends on endogenous properties of the company. This is typically the case for a system composed by quasi-uncoupled variables evolving over time and subject to random perturbations with a well-defined perturbation mean (for instance, Gaussian perturbations). On the other hand, if a company has a large irreversibility variance, its response varies largely with the type of perturbation. This is usually the case either for uncoupled variables subject to random perturbations with fat tail distributions (for example, power law distributed perturbation strength), or for highly coupled variables. For instance, AA (Alcoa Inc.) has large irreversibility Score, however reasonably low irreversibility variance according to figure \ref{example}, hence the dynamics underlying the evolution of the price of this company have not dramatically changed over time (i.e., external perturbations approximately have the same impact). On the other hand, AIG (American International Group) shows relatively large irreversibility score and variance. Hence AIG, over the period 1998-2012, was affected differently by different financial perturbations. In particular, the perturbation that originated in 2008 has a qualitatively larger effect on the company's internal dynamics than other perturbations; this can also be seen in figure \ref{example}. 

\begin{figure}
\centering
\includegraphics[width=0.85\columnwidth]{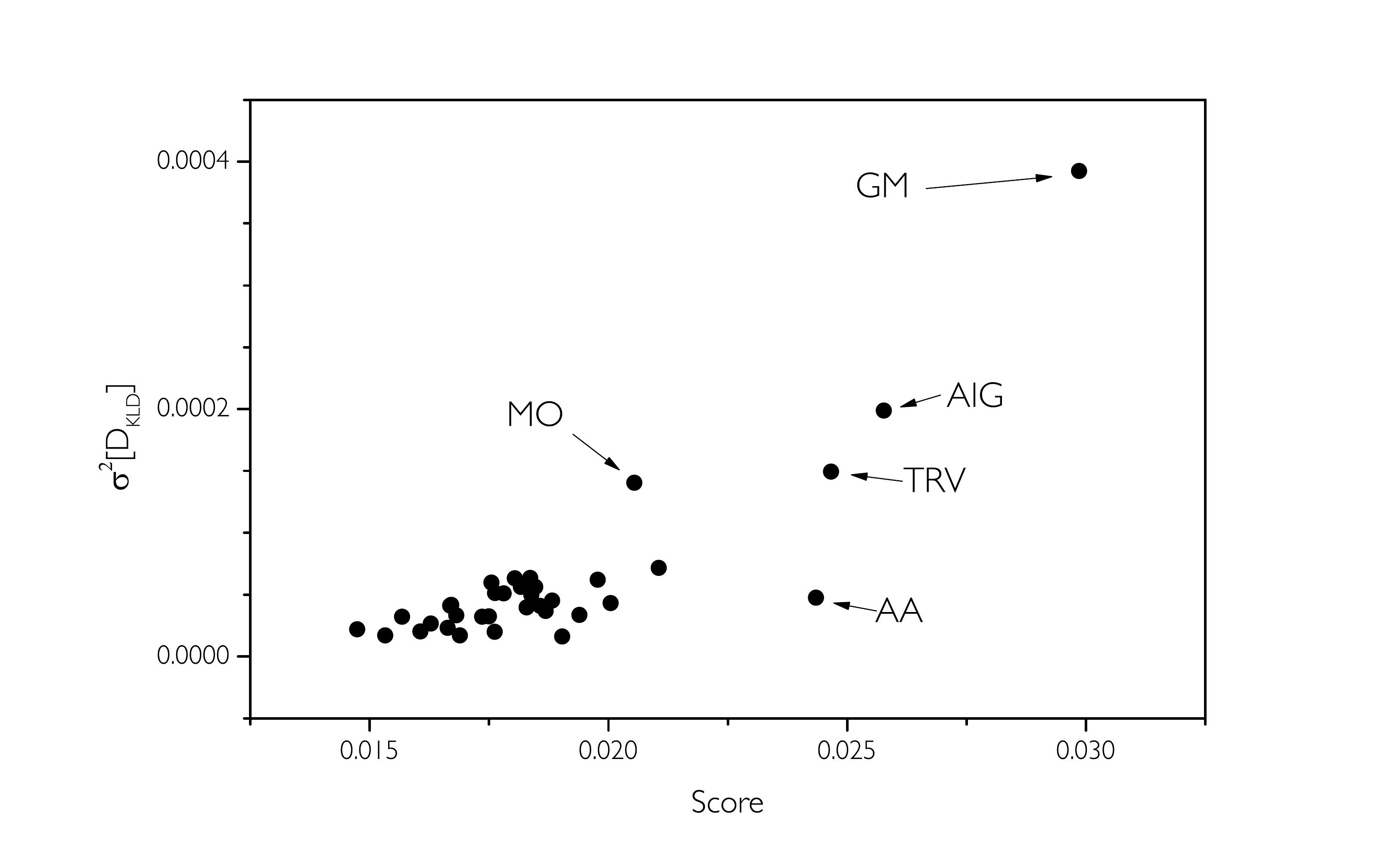}
\caption{Scatter plot of the irreversibility variance against the irreversibility score for each company.}
\label{scores}
\end{figure}

\noindent In general, the irreversibility Score will be a faithful static measure of a company's irreversibility as long as we have relatively small variance. We have compared these measures for all companies in figure \ref{scores}. We find a bulk of companies for which the Score ranges between 0.015 and 0.020, for which the variance is relatively constant. This means that the Score alone is a sufficient indicator of irreversibility, at least for these companies. Interestingly, we find that the top five multinationals in the Score ranking, also have large variance. These are companies which have been dramatically affected by major external perturbations at certain specific times, perhaps acting as global sensors of the  financial system's stability state. In the next section, we further explore this possibility, and investigate in an unsupervised way if the evolution of irreversibility features across companies over time reflects the stability of the whole financial system and thus allows us to classify and cluster periods of time according to their level of systemic reversibility. 

\noindent {\bf Assessing periods of financial reversibility.}\\
In order to be able to quantitatively compare the performance of the whole system amongst different periods of time, we consider the quantity $I_{\text{VG}}^c(\text{year})$, which is the irreversibility in company $c$ in the respective year, and create the vector $I_{\text{VG}}^{\bf c}(\text{year})$ where ${\bf c }=(c_1,\dots,c_{35})$ is the vector of companies.  %Analogously, element $IV_{\text{year}}(c_i)$ is the irreversibility variance of company $c_i$ over a certain year. 
Our entire database is then coarse-grained into a sets of 15 (the number of years), 35-dimensional observations (the companies) spanned by  $\{I_{\text{VG}}^{\bf c}(\text{year})\}_{\text{year}=1998}^{2012}$. In order to find patterns arising among different periods, we make use of two standard techniques in data mining: principal component analysis (PCA) and hierarchical clustering. 

\begin{figure}
%\centering
\includegraphics[width=0.48\columnwidth]{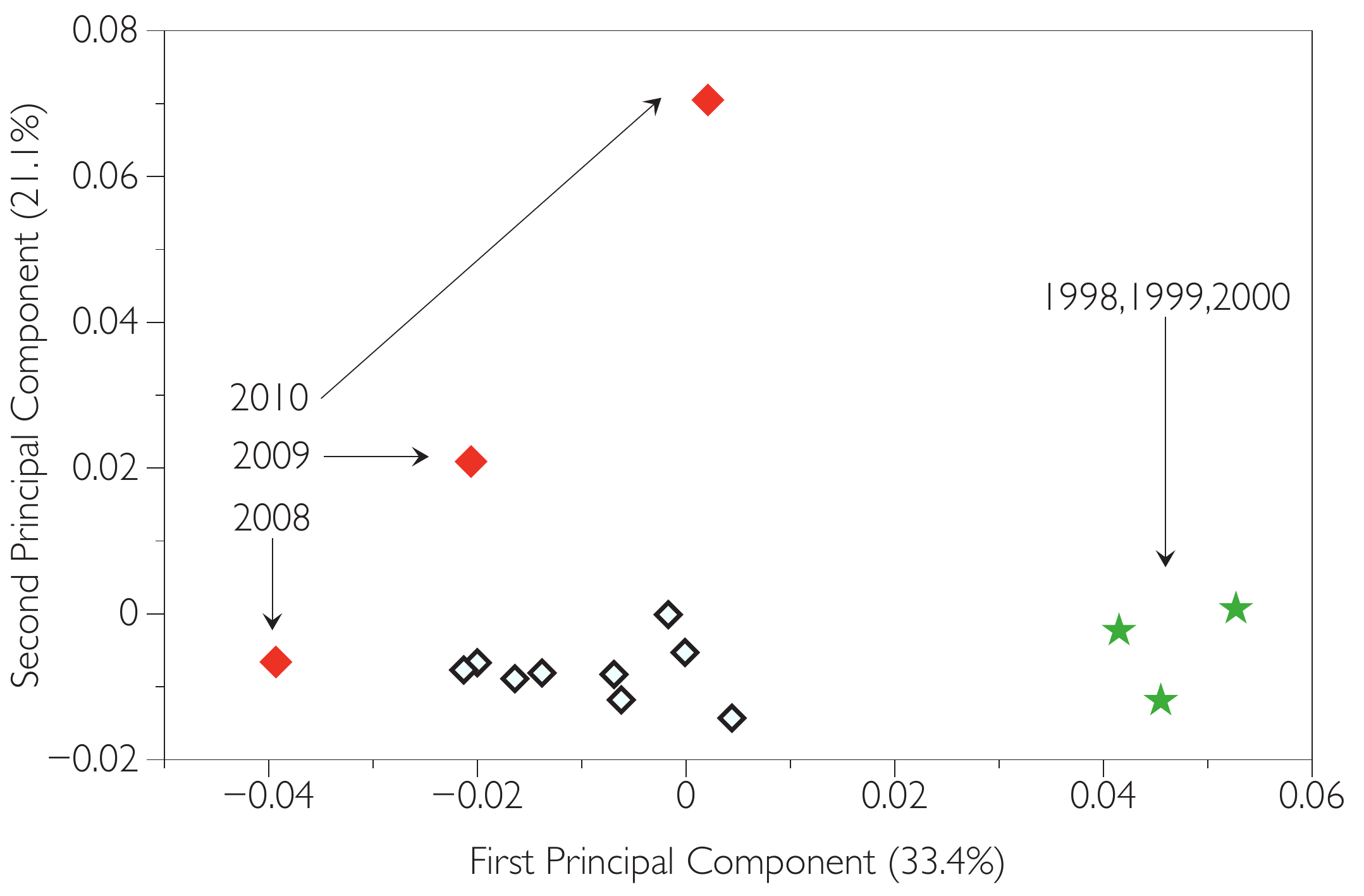}
\includegraphics[width=0.48\columnwidth]{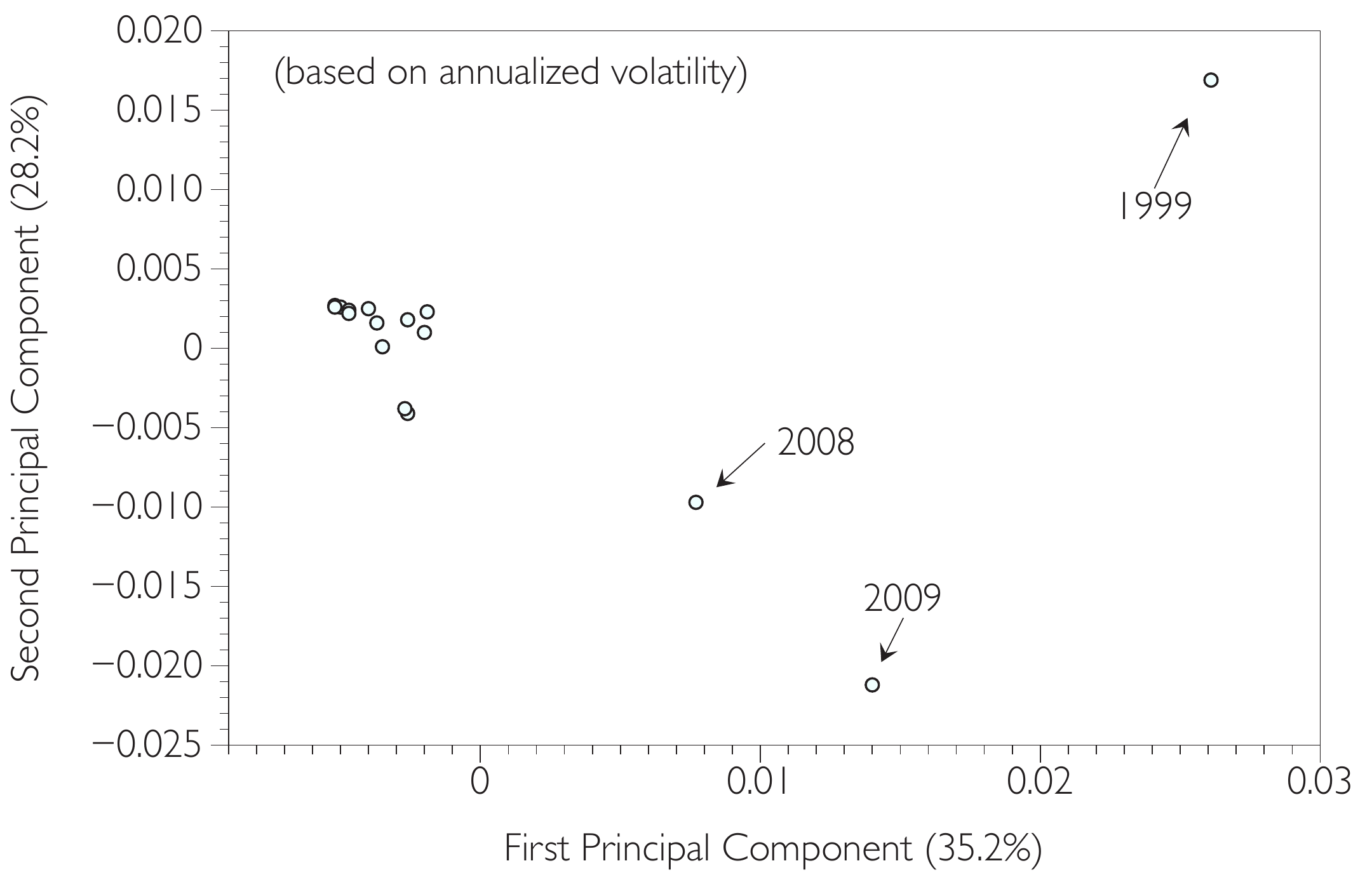}
\caption{(Left panel) Projection of financial periods in the PCA space of $\{I_{\text{VG}}^{\bf c}(\text{year})\}_{\text{year}=1998}^{2012}$. The first two principal components account for about $55\%$ of the system's variability. In the two dimensional space spanned by these components, we can already find three clusters, which account for somewhat stable years, dot-com bubble and the maximum of the global financial crisis. (Right panel) Similar analysis but using annualized volatilities. While results show some qualitative agreement, the analysis based on irreversibility provides a clearer picture.}
\label{PCAmean}
\end{figure}

\noindent Principal Component Analysis (PCA) \cite{PCA} is a common statistical procedure to perform dimensionality reduction on data. It uses an orthogonal transformation to project our set of observations, originally described in $\mathbb{R}^{35}$ -where each direction is possibly correlated among observations, as similar companies might have correlated irreversibility evolutions- into a lower dimensional subspace spanned by the so called principal components, obtained from the eigenvectors of the dataset covariance matrix. These particular directions are such that (i) they are orthogonal, (ii) the first principal component has the largest possible variance (that is, accounts for as much of the variability in the data as possible), and each succeeding component in turn has the highest variance possible under the constraint that it is orthogonal to (i.e., uncorrelated with) the preceding components. Thus
projecting each observation $O_i$ (originally $O_i \in \mathbb{R}^{35}$) into a smaller space spanned by the first $m$ principal components hugely reduces the dimensionality of the observations, while keeping the relevant information of the data. This projection is indeed the one that minimizes the mean squared distance between the data points and their projections. 

\noindent In the left panel of figure \ref{PCAmean} we show the projection of $\{I_{\text{VG}}^{\bf c}(\text{year})\}_{\text{year}=1998}^{2012}$ in the PCA space spanned by the first two principal components (accounting for about $55\%$ of the data variability). Interestingly, observations (i.e., years) automatically seem to cluster into three separated groups. The first one includes the observations for years 1998, 1999 and 2000 -a period that can be identified with the dot-com bubble-. The second group includes the years 2008-2010, which is in turn well known to represent the period of largest financial stress resulting from the global financial crisis. The third group amalgamates the rest of the years, but it is difficult to find finer structures within that one with this representation. Note that these results are on good agreement with recent, alternative metrics that make use of multiplex visibility graph mutual information \cite{multivariate}.

\noindent For the sake of comparison, in the right panel of figure \ref{PCAmean} we have performed a similar analysis, but using the annualized volatilities of each company as the features of the vectors characterizing each year, instead of the average irreversibilities. First, note that while irreversibility and volatility are not correlated (see figure \ref{scores2}), we obtain qualitatively equivalent results in PCA space. This suggests that both measures provide complementary information. However, by only using volatilities we seem to be losing detail as both financially unstable periods are underrepresented in this latter case.

\begin{figure}
\centering
\includegraphics[width=0.65\columnwidth]{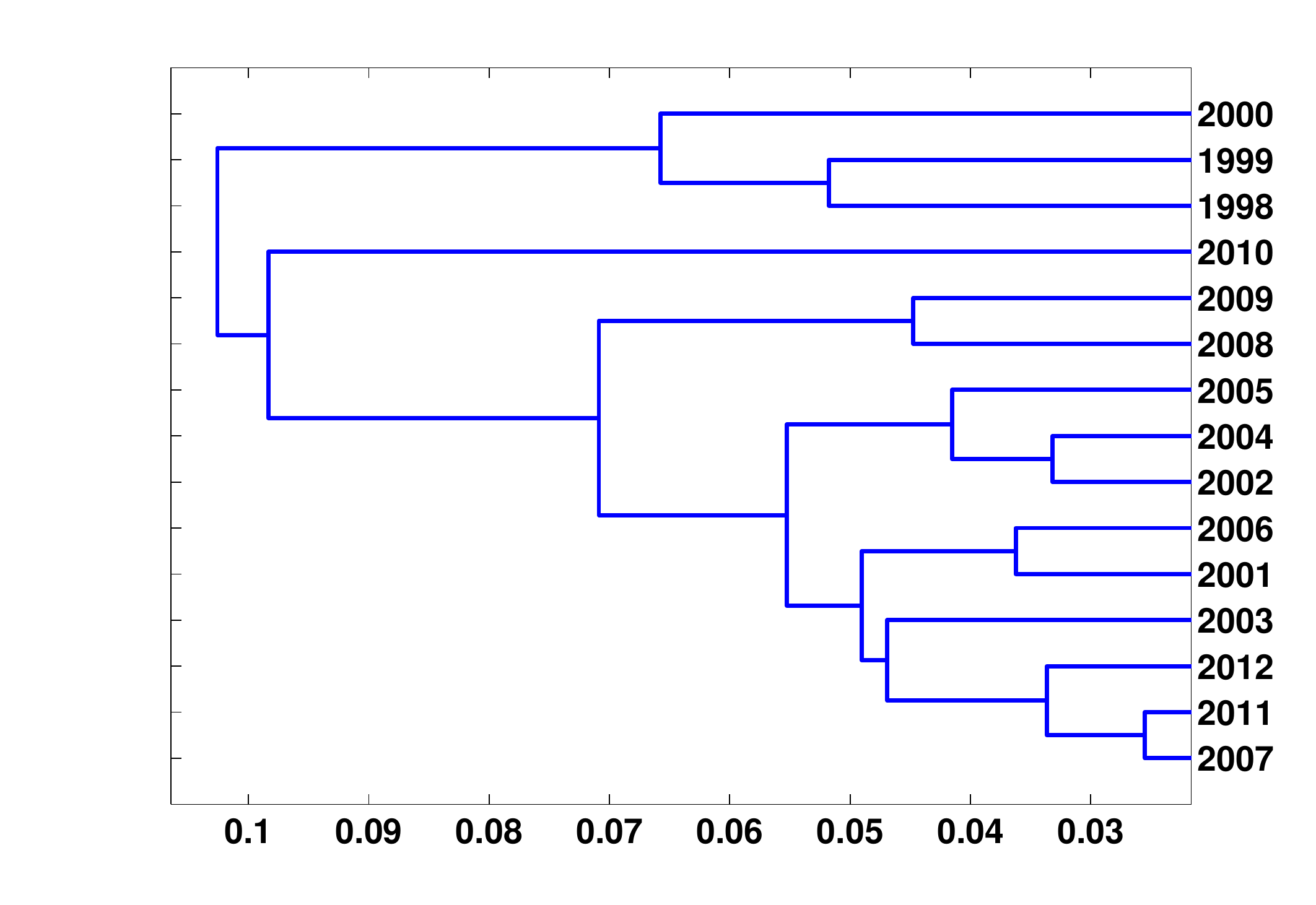}
\caption{Hierarchical clustering obtained from $I_{\text{VG}}^{\bf c}(\text{year})$. Years 1998, 1999 and 2000 group together at the top (analogous to the cluster seen in the left panel of \ref{PCAmean}, which can be interpreted as the dot-com bubble). Years 2008 and 2009 also group together.}
\label{HCA}
\end{figure}

\noindent 
Since the two dimensional space spanned by these projections only account for $55\%$ of the data variability, it is not totally straightforward that this particular two dimensional projection is \textit{faithful}, that is to say, we need additional evidence to confirm that any emerging separation or clustering is not spurious. In order to further explore in more detail the fine grained relation between years, we then make use of hierarchical clustering. This is a method of cluster analysis which seeks to build a hierarchy of clusters, where the merges and splits are determined in a greedy manner. The results of hierarchical clustering are usually presented in a dendrogram. To build this, we first compute a distance matrix $d$ among observations, where $d_{ij}$ is in this case the $L_2$ (Euclidean) distance between observations $i$ and $j$ in the original high-dimensional space $\mathbb{R}^{35}$: 
$$d_{ij}=\sqrt{\sum_{k=1}^{35} \bigg( I_i(c_k) - I_j(c_k) \bigg)^2}$$

\noindent A complete linkage criterion is considered to generate an agglomerative hierarchical cluster tree from the distance matrix $d$. In figure \ref{HCA} we plot the dendrogram generated via hierarchical clustering based on the $I_{\text{VG}}^{\bf c}(\text{year})$. This provides more visibly refined information compared to projection in PCA space. When we move left in the dendrogram, we are coarse-graining details.
For instance, we can see that years 1998, 1999 and 2000 group together (analogous to the cluster seen in the left panel of \ref{PCAmean}, which can be interpreted as the dot-com bubble). At the same level of granularity, 2008 and 2009 group together (which falls in the period of the global financial crisis). As we decrease the granularity, 2010 pairs up with 2008-2009, leaving a bulk of other years which can be identified with more stable and efficient financial periods 2001-2007 group together.  Interestingly, if we go at smaller scales, we also find that years 2011 and 2007 group together, suggesting that the onset and exit of the global financial crisis leaves a similar fingerprint on the irreversibility metric. All in all, these results are in good agreement with those found using PCA, and suggest that the evolution of the system's financial stability can be extracted from the collective evolution of companies irreversibility features.

\section{Discussion}
In this work we have extended and studied the concept of time irreversibility to the context of financial time series. While this is a statistical concept which can be linked with entropy production in (non-equilibrium) stationary states, it can be extended to the realm of \textit{non-stationary} time series by using the visibility algorithms. These transform time series into graph-theoretical representations and allow for a direct quantification of time series irreversibility even if the associated dynamical process is non-stationary.  

\noindent We found that the stock prices of the companies in our dataset are indeed time irreversible, in the sense that finite-size irreversibility values are higher than those found for reversible null models. This is yet more evidence that exposes the inefficiency of financial systems, supporting the violation of the classical efficient market hypothesis. It is however important to note that different companies have distinct time evolving irreversibility patterns, and some display periods of quasi-reversibility, which implies that (i) some companies are \textit{more} irreversible than others, and (ii) the degree of reversibility of each company varies over time. According to (i), one can rank companies. As there is a conceptual link between predictability and efficiency, one can argue that reversible time series are less predictable than irreversible ones. In this sense the ranking of companies based on stock price irreversibility could provide relevant information for traders and optimal portfolio designs, and we have shown that this information differs to that gained from volatility measures. According to (ii), one can also rank periods of financial stability. Concretely, we found that periods of financial turmoil, such as the dot-com bubble or the global financial crisis, can be easily identified and distinguished from periods of financial stability if we use the irreversibility values of each company as the features to feed clustering algorithms. 

\noindent We conclude that the concept of time irreversibility, adequately adapted to financial time series scenario via visibility algorithms, reveals complementary and valuable information on the evolution and the structure of stock prices. Further research needs to be done to assess in a quantifiable way the promising relation between irreversibility and predictability in this context. 

%%% List of companies %%%
\begin{table}
\begin{ruledtabular}
\begin{tabular}{cccc}
{\bf Acronym} &{\bf Name}& {\bf Comment}&{\bf Score Rank}\\
AA&Alcoa Inc.&NYSE&4\\
AIG&American International Group, Inc.&NYSE&2\\
AXP&American Express Company&NYSE&25\\
BA&The Boeing Company&NYSE&18\\
BAC&Bank of America Corporation&NYSE&28\\
C&Citigroup Inc.&NYSE&14\\
CAT&Caterpillar Inc.&NYSE&11\\
CSCO&Cisco Systems, Inc.& NYSE&24\\
CVX&Chevron Corporation&NYSE&9\\
DD&E.I. du Pont de Nemours and Company&NYSE&15\\
DIS&The Walt Disney Company &NYSE&31\\
GE&General Electric Company& NYSE&22\\
GM&General Motors Company&NYSE&1\\
HD&The Home Depot, Inc.&NYSE&8\\
HON&Honeywell International Inc.&NYSE&20\\
HPQ&Hewlett-Packard Company&NYSE&7\\
IBM&International Business Machines Corporation&NYSE&16\\
INTC&Intel Corporation&NasdaqGS&30\\
JNJ&Johnson $\&$ Johnson&NYSE&35\\
JPM&JPMorgan Chase $\&$ Co.&NYSE&21\\
KO&The Coca-Cola Company&NYSE&13\\
MCD&McDonald's Corp.&NYSE&12\\
MMM&3M Company&NYSE&32\\
MO&Altria Group Inc.&NYSE&6\\
MRK&Merck $\&$ Co. Inc.&NYSE&10\\
MSFT&Microsoft Corporation&NasdaqGS&34\\
PFE&Pfizer Inc.&NYSE&29\\
PG&The Procter $\&$ Gamble Company&NYSE&19\\
T&AT$\&$T, Inc.&NYSE&27\\
TRV&The Travelers Companies, Inc.&NYSE&3\\
UNH&UnitedHealth Group Incorporated &NYSE&5\\
UTX&United Technologies Corporation&NYSE&23\\
VZ&Verizon Communications Inc.&NYSE&17\\
WMT&Wal-Mart Stores Inc.&NYSE&33\\
XOM&Exxon Mobil Corporation&NYSE&26\\
\end{tabular}
\end{ruledtabular}
\caption{List of companies and associated irreversibility Score Rank (see the text).}
	\label{companies}
\end{table}

%\clearpage
\bibliography{apssamp}% Produces the bibliography via BibTeX.

\end{document}